\documentclass[nologo,11pt,a4paper,hidelinks]{ETHpaper}
\usepackage{graphicx, amsmath, amssymb,color,wasysym}
\usepackage[authoryear]{natbib}
\usepackage{caption}
\usepackage{subcaption}
\usepackage{longtable}
\usepackage{multirow}
\usepackage{booktabs}
\usepackage{mathtools}
\usepackage{cancel,cool}
\usepackage{enumitem}

\begin{document}

\newcommand{\mean}[1]{\left\langle #1 \right\rangle}
\newcommand{\abs}[1]{\left| #1 \right|}
\newcommand{\ul}[1]{\underline{#1}}
\renewcommand{\epsilon}{\varepsilon}
\newcommand{\eps}{\varepsilon}
\renewcommand*{\=}{{\kern0.1em=\kern0.1em}}
\renewcommand*{\-}{{\kern0.1em-\kern0.1em}}
\newcommand*{\+}{{\kern0.1em+\kern0.1em}}

\newcommand{\RA}{\Rightarrow}
\newcommand{\bbox}[1]{\mbox{\boldmath $#1$}}
\newcommand{\GC}[1]{{\color{red}#1}}

\renewcommand{\thefootnote}{ \fnsymbol{footnote} }

\title{Struggling with change: The fragile resilience of collectives}

\titlealternative{Struggling with change:  The fragile resilience of collectives}

\author{Frank Schweitzer,$^{1,}$\footnote{Corresponding author, \url{fschweitzer@ethz.ch}} Christian Zingg,$^{1}$ Giona Casiraghi,$^{1}$ }

\authoralternative{F. Schweitzer, C. Zingg, G. Casiraghi}

\address{$^{1}$ Chair of Systems Design, ETH Zurich, Weinbergstrasse 58, 8092 Zurich, Switzerland
}

\reference{(Submitted for publication)}

\www{\url{http://www.sg.ethz.ch}}

\makeframing
\maketitle

\begin{abstract}

Collectives form non-equilibrium social structures characterised by a volatile dynamics.
  Individuals join or leave.
  Social relations change quickly. 
  Therefore, differently from engineered or ecological systems, a resilient reference state cannot be defined. 
    We propose a novel resilience measure combining two dimensions: robustness and adaptivity.
  We demonstrate how they can be quantified using data from a software developer collective.
  Our analysis reveals a resilience life cycle, i.e., stages of increasing resilience are followed by stages of decreasing resilience.
  We explain the reasons for these observed dynamics and provide 
  a formal model to reproduce them.
  The resilience life cycle allows distinguishing between short-term resilience, given by a sequence of resilient states, and long-term resilience, which requires collectives to survive through different cycles. 

  \emph{Keywords: } Resilience, Intragroup Processes, Social Psychology, Quantitative Methods 
  \end{abstract}
\date{\today}

\section{Introduction}
\label{sec:Introduction}

When the CEO of a major Swiss telecommunication provider was asked about the long-term goal of his company, he replied: Still being in the market in five years.
This example could well serve as a shorthand description of resilience.
Being there in five years means that the company could either withstand shocks or recover from them if they could not be avoided.
For economic entities like communication firms, shocks could result from various sources, for instance, market disruptions from new competitors, legal regulations about privacy from governments, technological innovations that change communication behaviour, and many more.
Such incidents are likely to happen over time.
What makes them \emph{shocks} is their unpredictability.
Hence, a responsible CEO would probably prepare his company to cope with the unforeseeable; he would strengthen the company's ability to adapt to any changes quickly.

A similar understanding of resilience applies to individuals that face various mental, health, economic or social challenges.
In a mechanical sense, it would be difficult to define a ``stable state'' for them.
Their stability is indicated by the fact that they can master these challenges and still are ``there'', despite a very demanding life.
Instead of individuals, in the following, we focus on \emph{collectives}, i.e. social organisations comprising a larger number of individuals.
This term refers to formal or informal groups of interrelated individuals who pursue a collective goal and are embedded into an environment~\citep{Ostrom2009,Hoegl2001}.  Examples range from companies and non-profit organisations to high school classes or virtual teams collaborating via online systems.
We use as our running example a collective of developers of the open source software project \textsc{Gentoo} (for the details, see Appendix~\ref{sec:gentoo-data}).

Compared to other types of systems, e.g., technical or ecological systems, we know very little about the resilience of collectives, which provides the primary motivation for our paper.
We argue that the difficulties of tackling the resilience of collectives with a formal approach result from two \emph{dynamical} problems, discussed in the following.
The first one is the fast and continuing change within collectives, and the second one is the additional feedback cycle resulting from their response to changes induced by themselves.

\begin{figure}[t]\centering
  	\begin{subfigure}{.5\textwidth}\centering
  		\includegraphics[width=0.8\textwidth]{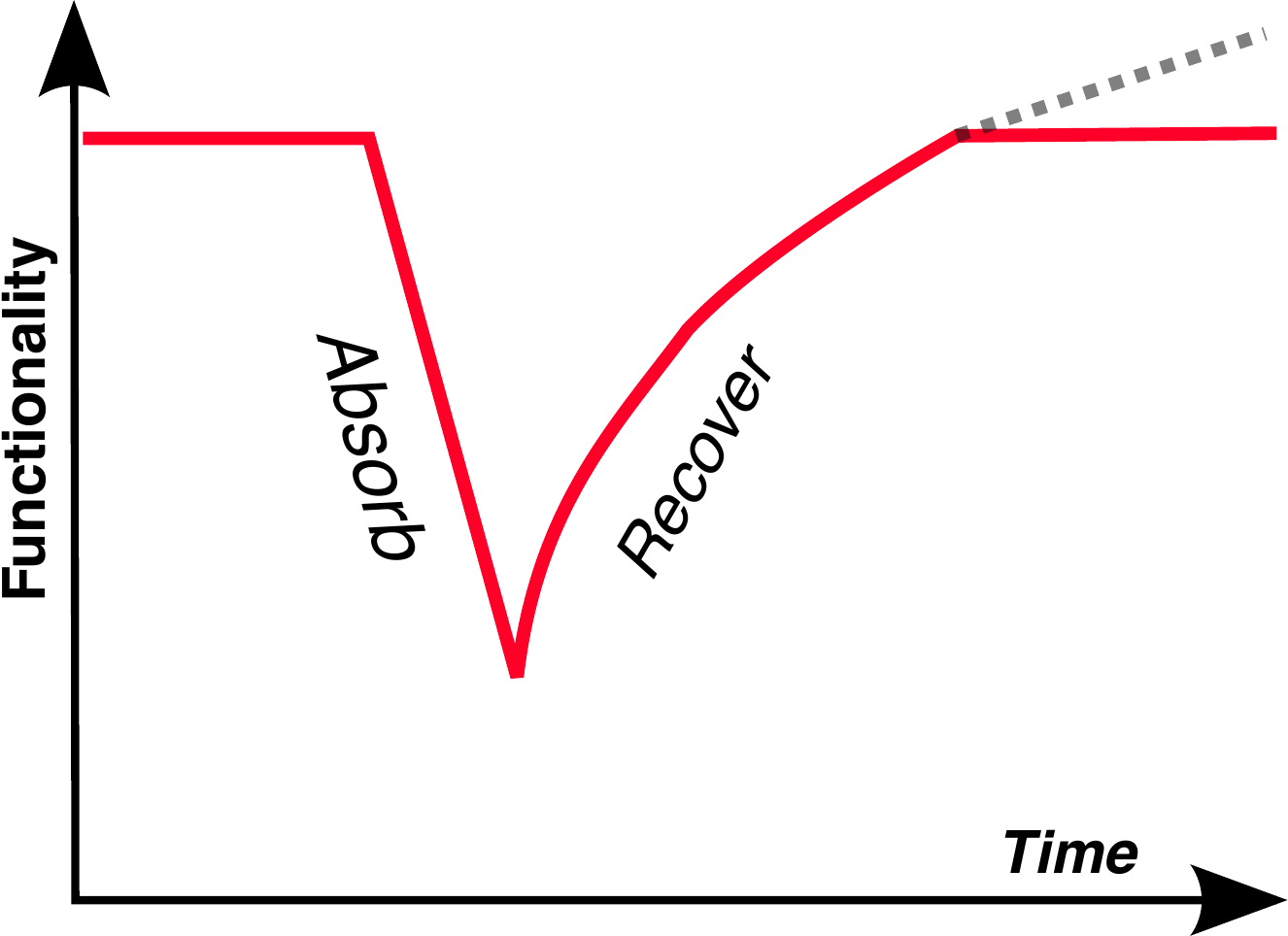}
  		\caption{}\label{fig:resilience:a}
  	\end{subfigure}\hfill
  	\begin{subfigure}{.5\textwidth}\centering
		\includegraphics[width=0.9\textwidth]{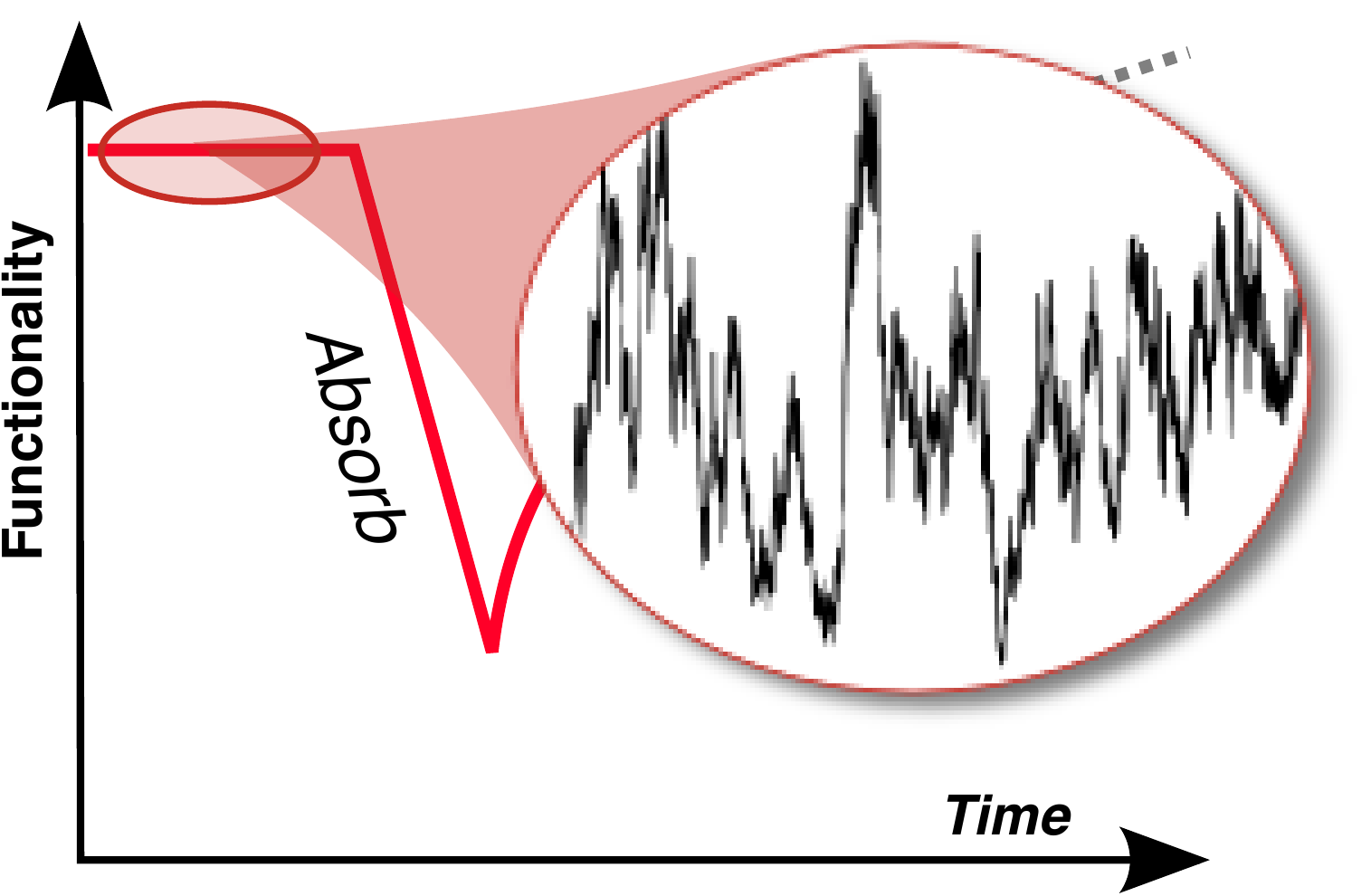}
  		\caption{}\label{fig:resilience:b}
  	\end{subfigure}
  \caption{Problems defining a reference state for resilience, understood as the ability to absorb shocks and to recover: (a) engineered system (e.g. a bridge), (b) social system (e.g. a collective).}
  \label{fig:resilience}
\end{figure}

Most collectives have in common that they are very \emph{volatile}.
They may experience fast changes in their structure, e.g. the number of individuals and their relations, have to cope with fluctuating task volumes or frequent interruptions, constant environmental impacts, etc.
This volatility makes them different from, e.g., engineered systems, which are built to last.
The common notion of resilience for engineered artefacts, such as bridges, is illustrated in Figure~\ref{fig:resilience:a}.
A bridge is planned for a defined functionality, e.g. a given number of cars per hour passing the bridge.
This functionality remains as long as no critical shocks appear, either caused by internal malfunction, e.g., lack of maintenance, or external disruptions, e.g., an earthquake.
If the shock happens, the bridge's functionality is partially or entirely destroyed.
Nonetheless, the bridge can be rebuilt, recovering the functionality and often even improving it.

The assumption underlying Figure~\ref{fig:resilience:a} is a known reference state, i.e. the planned functionality, which remains relevant over time.
For highly volatile systems, shown in Figure~\ref{fig:resilience:b}, we cannot define such a reference state, partly because it is hardly quantifiable and partly because it is constantly changing.
This implies that we are also unable to specify what we mean by a ``shock''.
Unlike the bridge, where shocks result in a measurable dropdown of functionality, we always have ``shocks'' of varying sizes.
The ability to recover is not restricted to the aftermath of a breakdown.
Instead, it requires a continuous effort from the collectives to adapt to all sorts of challenges.
Most importantly, the recovery is not an external intervention, like the repair of a bridge, but the result of an internal response of the collectives.
Consequently, we need a new and dynamical approach to the resilience of such social systems.

\section{A new resilience  measure}
\label{sec:resilience}

\subsection{Defining robustness and adaptivity}
\label{sec:robustn-adapt}

Resilience concepts have been developed in different disciplines, ranging from ecology to engineering, the social sciences, management sciences, or mathematics \citep{Hosseini2016}.
Its precise meaning differs across and sometimes even within these disciplines \citep{Fraccascia2018a,Baggio2015}.
Many approaches take resilience simply as a synonym for stability.
In ecology, for example,
a system is said to be resilient if, after a perturbation, it returns to a previously assumed stable state \citep{Grodzinski1990,Gunderson2000}.  This idea borrows from classical mechanics and thermodynamics with their definitions of equilibrium states as minima of some potential energy.

Collectives are inherently open non-equilibrium social systems.
Stationary states in non-equilibrium can only be kept if they are constantly \emph{maintained}, and collectives are no exception.
Their resilient state has to be actively managed.
Otherwise, it dissolves over time like any other ordered state.
So, what precisely has to be maintained?
We propose that resilience $\mathcal{R}[A(t),R(t)]$ is composed of a structural component that captures the \emph{robustness}, $R(t)$, and a dynamic component that captures the \emph{adaptivity}, $A(t)$, of a system, which both can change over time.

\begin{figure}[htbp]
  \centering
  \includegraphics[width=0.45\textwidth]{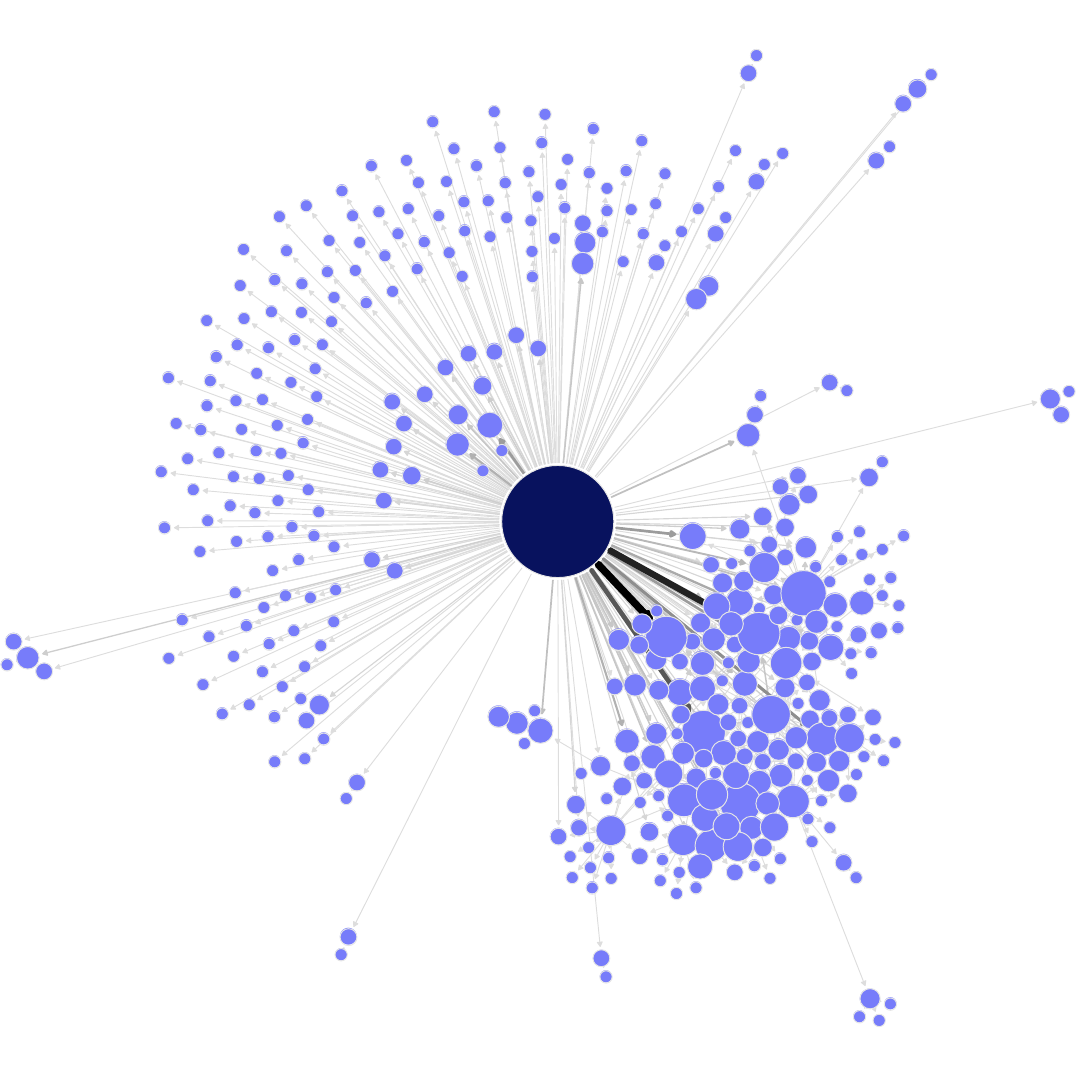}
  \caption{
    Network of task assignments between Gentoo developers in September 2007.
    A node's size and colour intensity is proportional to its degree.  }
  \label{fig:gentoo-network}
\end{figure}
Collectives can only function if they build on social structures.
In the example of a software developers' team, these structures are reflected by their work relations, communication channels, etc.
These structural features can be represented by a \emph{social network}, conveniently extracted from the project repositories using state-of-the-art tools such as \texttt{git2net} \citep{Gote2019}.
Links in this network are timestamped, directed, and weighted \citep{gote2021analysing}, and multiple relationships can be captured by multi-edge \citep{casiraghi2017relational} and multi-layer networks \citep{garas2016interconnected}.
This social network evolves if nodes or links are added or deleted, or links are rewired.
Collectives utilise this social structure for their activities.
A well-maintained social network will allow developers to, e.g., write more code, fix bugs faster, and reduce coordination overhead.

While robustness has an intuitive interpretation, adaptivity is more challenging to grasp.
It refers to the \emph{ability} of the collective to attain different states, not necessarily to actual transformations \citep{schweitzer2021fragile}.
Hence, adaptivity depends on the available \emph{options} to change the current state.
One way to measure this ability is \emph{potentiality} \citep{zingg2019entropy}, which quantifies how many different states become potentially available in a given situation.
This strongly depends on existing constraints for the collective.
The generalised hypergeometric ensemble of graphs (gHypEG) \citep{casiraghi2021configuration} allows calculating these states from an analytic approach.
Obviously, a collective cannot be resilient if it has no options to escape from an impaired situation.
Hence, the \emph{ability} to change is crucial for resilience.
However, knowing how the collective precisely evolves would imply predicting the future, which is not the aim of our approach.
In Appendix~\ref{sec:gentoo-data}, we summarise how the two variables, robustness and adaptivity, are operationalised using data from the software developers' collective.

\subsection{Composing resilience from robustness and adaptivity}
\label{sec:comp-resil-from}

\begin{figure}[htbp]\centering
	\begin{subfigure}{.5\textwidth}\centering
  		\includegraphics[height=0.75\textwidth]{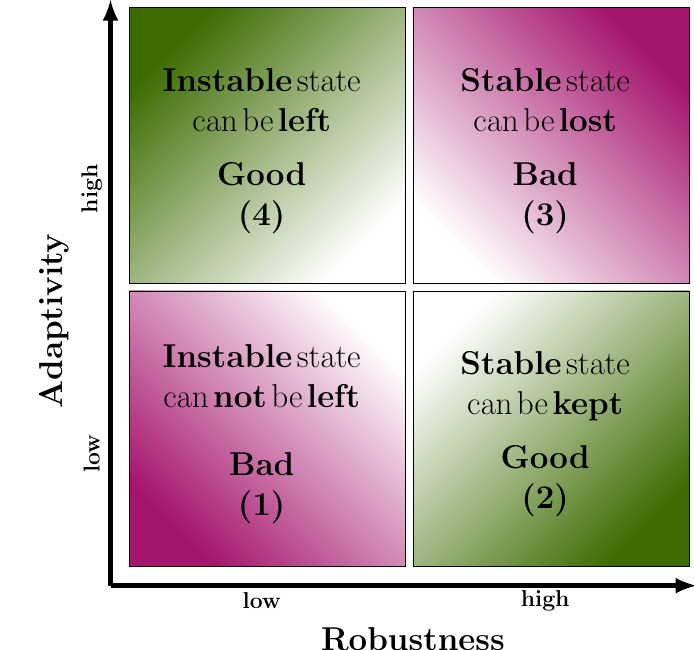}
  		\caption{}\label{fig:square:a}
  	\end{subfigure}\hfill
  	\begin{subfigure}{.5\textwidth}\centering
  		\includegraphics[height=0.75\textwidth]{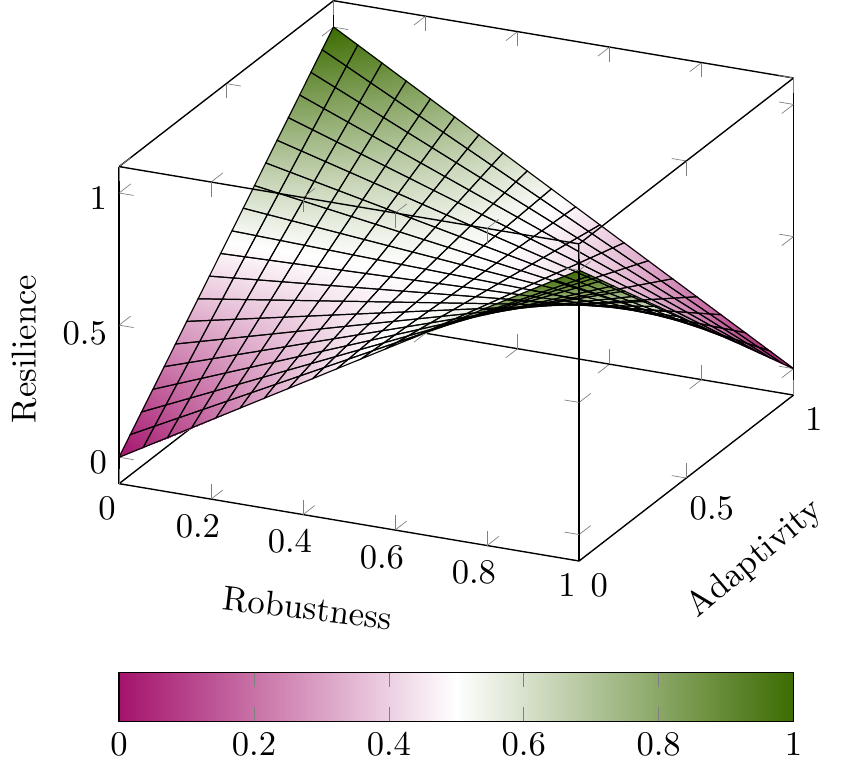}
  		\caption{}\label{fig:square:b}
  	\end{subfigure}
  \caption{ Resilience $\mathcal{R}$ as a function of robustness $R$ and adaptivity $A$: (a) Qualitative assessment of different states. (b) Quantification using Eqn.~\eqref{eq:5} with $A_{\mathrm{max}}=R_{\mathrm{max}}=1$ for illustration.}
  \label{fig:square}
\end{figure}

How does the resilience of collectives depend on their robustness and their adaptivity?
Should it be monotonously increasing with these two variables, assuming the more, the better?
The relations are more intricate, as we summarise in Figure~\ref{fig:square:a}.

\textbf{Region~(1)} is characterised by a low resilience of collectives because both robustness and adaptivity are \emph{low}.
Hence, there is nothing to build on, and the collective has few alternatives to change this unfavourable situation.

\textbf{Region~(2)} is characterised by high robustness, which implies a solid, structured social network.
It cannot be easily destroyed but also not be changed.
This state might be beneficial only if collectives \emph{should not} change because they are already close to an optimal state.
Collectives with high robustness have a lot to lose.
Thus adaptivity should be low to keep this state.
Only then can resilience become high.

\textbf{Region~(3)} is also characterised by high robustness, but the high adaptivity increases the risk that the collective could lose its robustness.
Therefore, such states have low resilience.
In the complementary case, if adaptivity should be \emph{high} because the collective needs different options to adapt, high robustness would only work against the necessary change.
Again, this means a lower resilience.

\textbf{Region~(4)} is characterised by low robustness.
That means the collective has nothing to lose, and alternative states will be better.
Thus, a high adaptivity can only improve the situation.
Therefore, resilience should be high.

Resilience, as a quantitative measure, should try to balance the influence of both robustness and adaptivity depicted in Figure~\ref{fig:square:a}.
This can be achieved quantifying resilience as proposed in \citep{IChingpaper} and shown in Figure~\ref{fig:square:b}:
\begin{align}
  \label{eq:5}
  \mathcal{{R}}(A,R)=R (A_{\mathrm{max}}-A)+A(R_{\mathrm{max}}-R)
  \end{align}

To summarise our discussion, adaptivity as the dynamic component of resilience is a two-edged sword.
It bears the chance to improve the bad states of collectives with low robustness and the risk of destroying good states with high robustness.
We also note that robustness or adaptivity \emph{alone} cannot guarantee that a collective is resilient.
Unlike robustness, which describes the current state, resilience has to reflect the ability to improve in the near \emph{future}.
Conversely, without the ability to adapt, collectives can be stable or unstable, but they are not resilient, i.e., they cannot respond to internal or external changes.

\subsection{A formal model to build up resilience}
\label{sec:formal-model}

We now proceed in two directions.
First, we study a formal model of generating resilience from robustness and adaptivity.
This will result in hypotheses for the behaviour of collectives.
Secondly, we test these hypotheses using data from our team of software developers.

From the above discussion, it becomes clear that robustness has to lead the improvement of the resilience of collectives, as all further activities depend on the existing social network.
At the same time, maintaining the social network also requires adaptivity.
New nodes have to be integrated.
Links have to be rewired or reinforced.
Therefore, the dynamics of robustness $R$ and adaptivity $A$ are coupled in a nonlinear manner.
For the details see Appendix \ref{sec:formal-relations}.

\begin{figure}[htbp]
  \centering
  \includegraphics[width=0.45\textwidth]{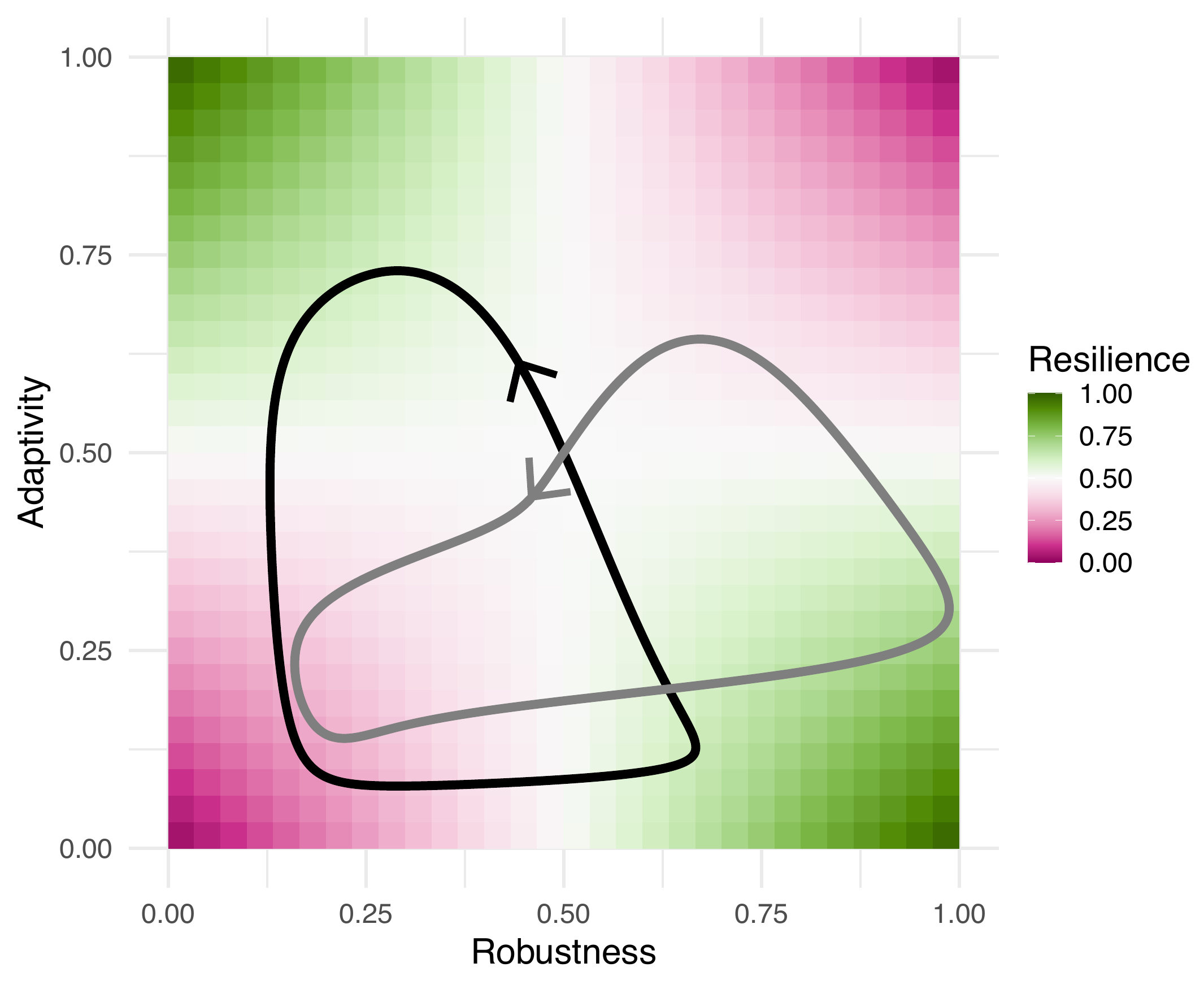}
\caption{Resilience trajectory in the phase space of robustness and adaptivity.
    The two scenarios (grey: I, black: II) are obtained from the formal model presented in Appendix~\ref{sec:formal-relations} for two different parameter sets. The color code refers to the Regions defined in Figure~\ref{fig:square:a}.}
  \label{fig:loop}
\end{figure}

As Figure~\ref{fig:loop} shows, the formal model generates distinct trajectories in the phase space of robustness and adaptivity for collectives' dynamics.
They resemble cycles, i.e., \emph{life cycles} in the development of collectives.
We show two different trajectories starting in Region (1) of low resilience, characterized by low robustness and low adaptivity.
The trajectories then quickly turn towards Region (2) of high resilience, characterized by high robustness, while adaptivity is low enough not to destroy the resilient state.
This region would be fortunate for the collective if it could stay there.
This, however, is not the case.
Our model predicts two scenarios shown in Figure~\ref{fig:loop}, which will be compared to the software developers' data.
Starting from Region (2), in Scenario (I), robustness remains high, but adaptivity is further increased such that Region (3) is reached.
In this region, resilience is low because the robust social structure is at risk of being lost because of too many alternative states and too little attention to maintain the current one.
Consequently, a failure follows, and the trajectory returns to the initial Region (1), where both robustness and adaptivity are low.
There, a new life cycle could start.

In Scenario (II), starting from Region (2), robustness decreases at the expense of adaptivity, which increases, such that Region (4) is reached.
Adaptivity and robustness are both coupled and, for certain parameter regions, cannot be increased simultaneously.
Such a coupling first leads the collective to another state of high resilience in which robustness does not work against adaptivity.
However, this state cannot be kept for long because robustness, the precondition of adaptivity, is low.
Therefore, after adaptivity has decreased, a failure follows, and a new cycle can start from Region (1).

These two scenarios are different in their sequence of resilient ($\square$) and nonresilient ($\triangle$) states.
Scenario (I) follows $(\triangle)\to (\square)\to (\triangle)\to (\triangle)\to ...$,
while Scenario (II) follows $(\triangle)\to (\square)\to (\square)\to (\triangle)\to ...$.
In Section~\ref{sec:comparison}, we will test the two hypotheses about the life cycle against the data from the developer collective and discuss the reasons for the collective failure in more detail.

\section{Resilience at work: An application}
\label{sec:comparison}

\subsection{Measuring resilience for a collective }
\label{sec:meas-resil-coll}

To demonstrate the applicability of our resilience measure, we analyse data from the \emph{bug handling collective} of \textsc{Gentoo} (see Appendix~\ref{sec:gentoo-data}).
Between October 2004 and March 2008, a central developer, named \emph{Alice} in the literature \citep{Zanetti2013,Garcia2013b}, became the most central figure in this collective (see also Figure~\ref{fig:gentoo-network}).
She assigned most bug reports to other developers for a while but left the project suddenly in March 2008.
The unforeseeable drop out of a core developer was a severe shock for the collective, which struggled for several years before it could reach a comparable level of operation.
\citet{Zanetti2013} already studied how different network measures reflect the dropout of Alice, while \citet{Casiraghi2021} developed a load redistribution model of task reassignments to study the likelihood of team failure.
For us, the recorded data allows studying the resilience of the collective during this period.

First, we constructed a social network from the available interaction data, where nodes indicate developers and directed links task assignments.
Because this network changes daily, we used a 30-day sliding window for aggregation.
Applying our quantitative measure for resilience requires operationalising the two main factors, robustness and adaptivity, for this network.
The details are again presented in Appendix~\ref{sec:gentoo-data}.
Robustness, as the structural component, is large if the nodes in the network have a similar degree.
That means everyone in the collective processes roughly the same number of tasks, either by solving or reassigning them,  and nobody gets overloaded.
Adaptivity, as the dynamic component, compares the number of developers assigning tasks to this number six months ago.
If it increases, more developers become potentially involved in bug handling.
Thus, the workload is better balanced, alternative members for task processing are available, and the time to process them becomes shorter \citep{Zanetti2013}.

\begin{figure}[htbp]
  \centering
\includegraphics[width=0.45\textwidth]{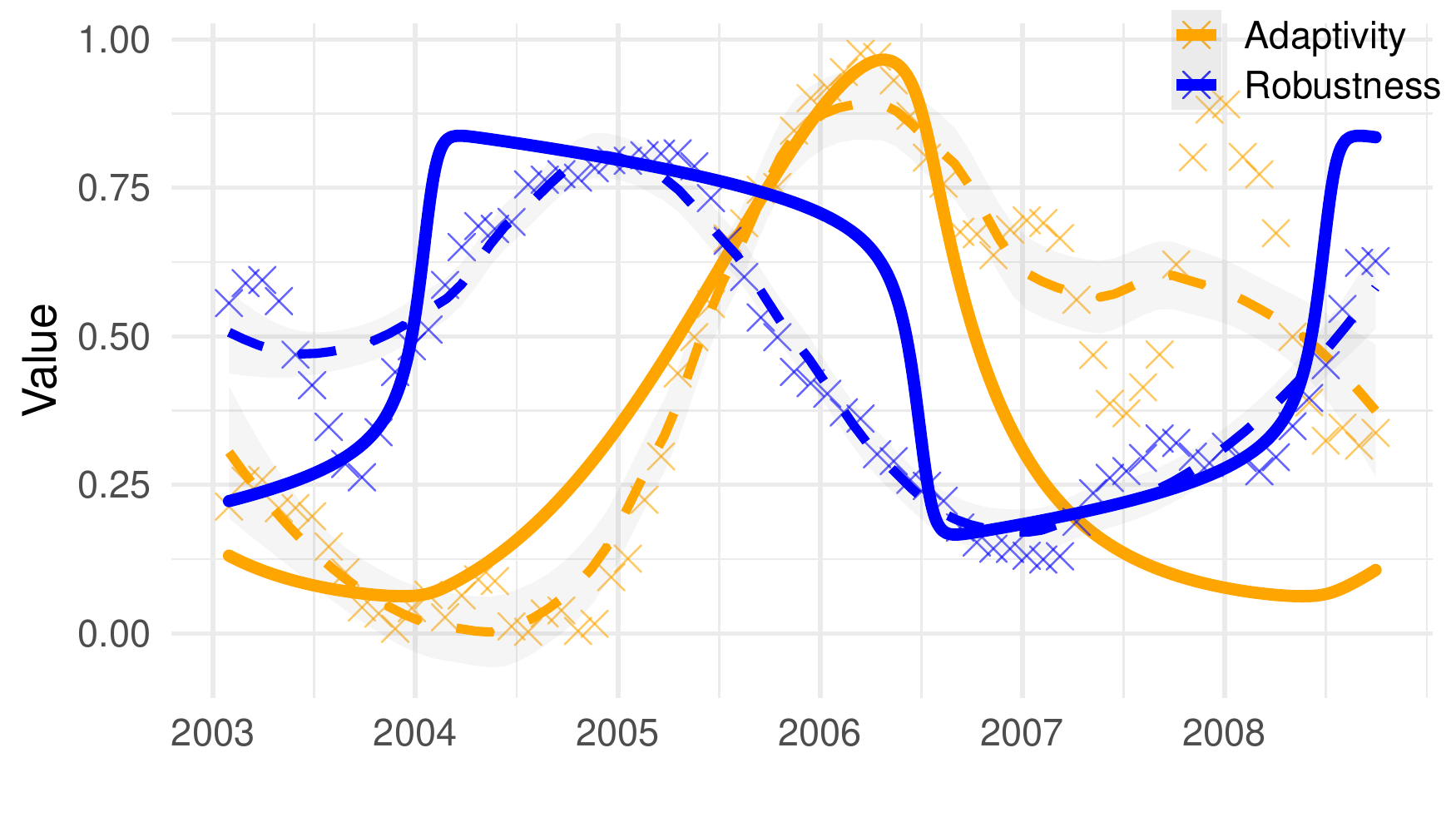}
  \caption{Robustness and adaptivity over time. Dots indicate the values obtained from the social network. Using a kernel density estimation, we reduce this information to the empirical curves, from which the fits to the dynamic model of robustness and adaptivity, Appendix~\ref{sec:formal-relations}, are obtained. }
  \label{fig:time}
\end{figure}

The results in Figure~\ref{fig:time} reveal the following scenario of how this collective copes with change.
Initially, adaptivity is low because the collective first has to establish a robust social structure for collaboration.
As this progresses, adaptivity also increases because more options become available for performing tasks.
In the same way, if robustness decreases, adaptivity follows the decrease with a time lag of several months.
That means robustness is instrumental for generating activity and ensuring adaptivity.
This is also reflected in our formal approach (Appendix~\ref{sec:formal-relations}).

Our attention shall focus on the time interval after 2004 when robustness started to decrease.
According to our operationalisation, this indicates that the task assignment became more centralised.
It was the time when developer Alice started to assign most of the tasks.
Interestingly, this concentration led to an increase in adaptivity, i.e. the number of developers who got tasks assigned still increased.
That means Alice effectively utilised the collective's workforce, involving more members.
However, the further concentration of the responsibilities eventually led to a decrease in adaptivity, i.e., fewer options for the collective to contribute.

\subsection{Explaining the failure}
\label{sec:explain}

The findings from our case study are remarkable in different aspects.
First, in Figures~\ref{fig:loop} and \ref{fig:time} we observe a \emph{life cycle}, i.e., the resilience of the collective first increases to decrease afterwards rapidly.
After returning to the initial low resilience state, the collective starts to consolidate again by building up robustness and adaptivity.
The dynamic model presented in Appendix~\ref{sec:formal-relations} is compatible with such a life cycle behaviour but, obviously, cannot capture additional perturbations occurring in this particular collective.
For example, in Figure~\ref{fig:time}, one can notice a deviation of adaptivity between the model and the data in 2007.
This was caused by the fact that during that period the central developer was temporarily suspended by the collective.
Such singular events cannot be reflected by our dynamic model.
But, interestingly, they do not change the principal dynamics of the life cycle.

Secondly, thanks to our dynamic resilience concept, we can understand the reasons behind the life cycle.
These are the adaptive processes inside the collective that push it out of the resilient state and eventually causes the failure. 
This reminds of self-organized criticality, a dynamic phenomenon in non-equilibrium complex systems and networks \citep{watkins201625,PhysRevE.85.026103}, where feedback processes drive a system into unstable states.
However, at difference to mechanical or physical systems, the dynamics approaching the critical state is not universal, but depends on the goal of the collective and the social mechanisms at work.

Specifically, the two states of high resilience for the collective are of different natures.
The state with high robustness in Region (2) is characterised by balanced interactions between developers, who were all involved similarly in assigning, redistributing and solving tasks.
However, little changes to the social network occurred because strategies to integrate new developers were missing.
Following the advent of Alice, the collective evolved to a second resilient state in Region (4).
In this state, adaptivity increased because more developers were involved in solving the tasks, and new members could be quickly and easily integrated into the organisational structure.
However, the effort to assign tasks became more centralised, and links that have become redundant disappeared from the social network.
Therefore robustness decreased.

This development reflects an internal reorganisation in the workflow.
With Alice as the central developer, the collective obtained a hierarchical organisation.
It became highly efficient regarding the task assignments but also highly vulnerable because the collective depended on a single individual.
Once the robustness has critically lowered, an adaptivity increase eventually destroys the previous resilient state.

The fact that the failure is caused \emph{endogenously}, i.e., by the collective itself, makes it different from engineered systems, such as the bridge example mentioned earlier, which may fail because of exogenous shocks.
It also makes it different from ecological systems, which often reach a stationary non-equilibrium state that persists over a long time if no critical perturbations occur.
Social systems constantly adapt to changes caused either exogenously or endogenously.
This response leads to intended as well as unintended consequences.
The intended one was the increased efficiency in utilising the workforce, thanks to the central developer.
The unintended one was the increased dependency on this central developer, causing the unnoticed erosion of robustness.

The life cycle observed allows us to characterize resilience in a more general manner.
Collectives could be seen as resilient \emph{only} if they follow more than one round of the life cycle.
This denotes a \emph{higher-order}, or \emph{long-term}, resilience.
A \emph{first-order}, or short-term resilience in contrast refers only to one cycle.
There, we already observe resilient states of the collective which can last for long, but are eventually destroyed by the adaptive dynamics.
Long-term resilience addresses the question how a collective is able to cope with a collapse.
The collective of the \textsc{Gentoo} developers was able to recover, albeit on a longer time scale that is not covered in our data set.
But other software development projects were not able to build up this long-term resilience, they disappeared after a few years.

\section{Discussion} \label{sec:comparison:-what-do}

\subsection{Comparison with existing approaches}
\label{sec:comp-with-exist}

Our analysis clarifies why existing resilience concepts cannot provide a comparable, quantifiable insight into the failure of the developers' collective.
They largely miss the coupling between structure and dynamics, expressed in the nonlinear relation between robustness and adaptivity.
Instead, they treat these dimensions as independent or, more often, only focus on robustness and stability.

Robustness models of networks are a prime example of such lopsided resilience concepts.
They can be classified into different approaches.
One group of models simulate attacks on the network structure by removing links or nodes.
Resilience is then measured as the size of the largest connected component surviving  \citep{Kitsak2018, casiraghi2020improving,schweitzer2020intervention}.
Another group of models simulate failure cascades after a shock initiates processes of load redistribution \citep{Burkholz2016b, Garcia2013, Cohen2000a}.
The size of failure cascades serves as a resilience measure.
Such attempts model only the robustness of the networks.
They leave out the ability of the network to respond, i.e. the adaptivity that is stressed in this paper.

While these models consider at least a time dimension for cascades and redistribution, other concepts simply take static topological network measures as proxies for resilience.
For instance, closeness centrality was applied as a resilience measure for transportation systems \citep{Ilbeigi2019} and infrastructure systems \citep{Omer2014}, but also for software developer teams
\citep{Zanetti2013}.
such measures only capture the structural dimension of resilience.

Adaptivity, which we have identified as the second dimension of resilience, is often discussed only as a synonym for dynamics.
Concepts such as first-passage times take explicitly into account only the time a system needs to return to a previously occupied equilibrium state after a perturbation \citep{Grodzinski1990}.
If a perturbation leads the system to transit to a different equilibrium state, this is known as \emph{robust adaption}, combining the notions of robustness and adaptation.
But adaptivity is reduced to a simple relaxation dynamic, whereas the volatility of collectives requires modelling a continuous dynamics.
This is often considered as \emph{adaptive capacity}.

It can be expressed in several different ways, for instance, in terms of the ability to learn and store knowledge, the ability to anticipate disruptive events, the level of creativity in problem-solving, or the dynamics of organisational structures~\citep{Folke2002,Smit2006}.
Some of these aspects have been assessed through survey research designs.
Examples are learning capability~\citep{Svetlik2007}, situational awareness, creativity~\citep{Mcmanus2007}, or the fluidity of structures~\citep{Goggins2014}.
The problem in measuring adaptive capacities is usually operationalisation.
Moreover, we miss a formal relation between adaptivity and robustness to understand resilience fully.

The literature further provides examples of more general resilience measures.
\citet{Hosseini2016} distinguishes the following categories:
\begin{description}[noitemsep,nolistsep]
\item (1) \emph{Conceptual frameworks} aim to find qualitative best practice recommendations to ensure a system's resilience.
\item   (2) \emph{Semi-quantitative indices} entail surveys asking experts to rate different resilience factors for a system on a scale, e.g. from $0$ to $10$.
\item   (3) \emph{General resilience measures} quantify the resilience of a specific class of systems, such as civil infrastructure or transportation systems.
\item   (4) \emph{Structural-based modelling approaches} model individual systems with domain-specific resilience factors.
\end{description}
This elaborate classification highlights that neither a universal resilience definition nor a measure working in all scenarios exists \citep{Carpenter2001, Meerow2016, Walker2012}.
Apart from this, the real problem behind most of these approaches is the \emph{quantification} of resilience factors and the efficiency in obtaining information.
We wish for measures that can be automatically and instantaneously calculated based on available data about collectives to monitor resilience continuously.
In contrast, almost every existing resilience measure is based on an \emph{ex-post evaluation}.
This approach may help us to understand why some failures have happened, but it is not sufficient to see them coming.

It is one of the main achievements of our framework that it allows precisely this: (i) quantification, (ii) monitoring, (iii) early warning in case of risky situations.
Moreover, the concepts of robustness and adaptivity underlying our resilience approach also allow a better understanding of the \emph{reasons} for decreasing resilience.

\subsection{A need for further research}
\label{sec:need-furth-rese}

The need for overarching, quantitative and explanatory resilience measures for collectives has been pointed out in the literature for long.
\citet{Davidson2010} emphasises that \textit{``the current [resilience theory] is not readily applicable to social systems.''}
She mentions reasons such as the ability of social systems to postpone the effects of disruptions, unequally distributed agency, humans' ability to anticipate risks, complex power relations, or a tendency for complex collective actions in social systems.
\citet{Al-Khudhairy2012} argue similarly, emphasising the ability of collectives to adapt and self-organise as essential building blocks of resilience.
They acknowledge that existing studies are \textit{``still at the very early stages to learn how to design resilient groups and organisations.''}

As we have demonstrated in our analysis, it is not sufficient to import existing measures or factor classifications from other disciplines to fill the research gap about the resilience of collectives.
They must not be uncritically applied to collectives because \textit{``human systems embody power relations and do not involve analogies of being self-regulating or `rational' ''} \citep{Cannon2010}.
Static resilience measures may work for engineering artefacts but not for volatile social systems where change is the new normal.
Hence, studying the resilience of collectives requires developing a \emph{dynamic} approach that reflects the non-equilibrium conditions and the permanent adaptivity.

But there is more to it.
In fact, resilience is a \emph{system specific} response to a \emph{specific shock}.
That means any approach to understanding resilience has to be contextualised with respect to specific collectives.
Operationalisations for collectives' robustness and adaptivity, thus, have to reflect the available data.
In this paper, we have used the example of a collective with the specific goal of collaboration, where the data allowed us to employ a social network perspective.
This is not always guaranteed.
Our framework of constituting resilience from robustness and adaptivity can rightly claim to provide a new and overarching perspective for collectives.
However, the specific measures for these two dimensions have to be developed with concrete collectives and concrete data in mind.
Ideally, such measures shall reflect the micro-processes generating social resilience.
In turn, this would open the door for mechanism design to improve resilience in collectives.

\subsection*{Acknowledgements}

The authors thank Ingo Scholtes, David Garcia, Antonios Garas and Pavlin Mavrodiev for discussions.

{\small \setlength{\bibsep}{1pt}

}

\begin{appendix}
  \section{Formal dynamic model of robustness and adaptivity}
\label{sec:formal-relations}

For convenience, we introduce reduced variables $r=\ln(R)$, $a=\ln (A)$, for which the dynamics are specified in the following.
Both robustness and adaptivity require to have a positive maintenance term.
On the other hand, both cannot grow infinitely but are bound to a maximum value, dependent on the system under consideration.
Therefore, we have to consider a negative decay term.
In the case of robustness, too much activity could destroy a resilient state.
Therefore, large values of $a$ should lead to a decrease in $r$.
Further, robustness can only be established and increased based on the existing structure.
Thus, it has a positive impact on its growth.
Also, adaptivity requires functionality and, therefore, a certain level of robustness.
These considerations lead to
\begin{align}
  \label{eq:1}
  \frac{dr}{dt}&=\alpha_{r}I_{r}(t)+\gamma_{r}r(t) -\beta_{a}a(t)\nonumber \\
  \frac{da}{dt}&=\alpha_{a}I_{a}(t)-\gamma_{a}a(t)+\beta_{r}r(t)
\end{align}
We assume that the effort per time unit is constant and shared between the maintenance of robustness and adaptivity using a model parameter $0<q<1$.
\begin{align}
  \label{eq:2}
  I_{r}(t) &= (1-q)\;;\quad   I_{a}(t) = q
\end{align}
Eventually, the impact of robustness on its further increase is not a constant but a nonlinear function of $r$,
$\gamma_{r}=\gamma_{r_{0}}-\gamma_{r_{2}}r^{2}$.
This assumption reflects that the positive impact of robustness is primarily important if no social relationships or established organizational structures exist yet and becomes less critical if already higher levels of robustness are obtained.
This leads to the coupled dynamics of robustness and adaptivity in the following form
\begin{align}
  \label{eq:4}
  \frac{dr}{dt}&= \alpha_{r}{(1-q)} +\gamma_{r_{0}}r -\gamma_{r_{2}}r^{3}-\beta_{a}a  \nonumber \\
  \frac{da}{dt}&=\alpha_{a} {q}- \gamma_{a}a+\beta_{r}r
\end{align}

\section{Gentoo Data}
\label{sec:gentoo-data}

\textsc{Gentoo} is a computer operating system based on the Linux kernel.
The developers who fix bugs in \textsc{Gentoo} use the software  \textsc{Bugzilla}~\citep{TheBugzillaTeam2018} to coordinate their work since 2002.
Such an extended use of bug-tracking software allows us to access their interactions in detail.
This paper uses the data set by \citet{Zanetti2013}.
The data set contains $45,086$ task assignments between $8,591$ developers from January 2003 to October 2008.

We compute the empirical robustness $\overline R$ and adaptivity $\overline A$ on adjacent 30-day windows.
In each window, the collective is represented as a network $g_t$.
Nodes correspond to developers and edges to their task assignments.
An example of the network is shown in Figure~\ref{fig:gentoo-network}.
For each time window, we calculate the normalised degree centralisation \citep{Wasserman1994}, $m_{t}$, of the  undirected largest connected component of $g_t$.
Robustness at time $t$ is then defined as the average of (1-$m_{t}$) over the sliding window.
Likewise, adaptivity at time $t$ is defined as the average change in the number of task assigners at time $t$ compared to half a year earlier.
To generate Figure~\ref{fig:time}, we ensure that the empirical measures  defined in the interval $[0,\infty)$ are compatible with our model.
Therefore, we apply a transformation mapping them to the interval $(0,1)$.

\end{appendix}

\end{document}